\documentstyle[12pt]{article}

\def\Msun{\ifmmode{~M_\odot}\else$M_\odot$~\fi}
\def\rsun{\ifmmode{~r_\odot}\else$r_\odot$~\fi}
\def\kms{\ifmmode{$~km\thinspace s$^{-1}~}\else km\thinspace s$^{-1}~$\fi}
\def\ga{\mathrel{\mathchoice {\vcenter{\offinterlineskip\halign{\hfil
$\displaystyle##$\hfil\cr>\cr\noalign{\vskip1.5pt}\sim\cr}}}
{\vcenter{\offinterlineskip\halign{\hfil$\textstyle##$\hfil\cr>\cr
\noalign{\vskip1.0pt}\sim\cr}}}
{\vcenter{\offinterlineskip\halign{\hfil$\scriptstyle##$\hfil\cr>\cr
\noalign{\vskip0.5pt}\sim\cr}}}
{\vcenter{\offinterlineskip\halign{\hfil$\scriptscriptstyle##$\hfil
\cr>\cr\noalign{\vskip0.5pt}\sim\cr}}}}}
\def\la{\mathrel{\mathchoice {\vcenter{\offinterlineskip\halign{\hfil
$\displaystyle##$\hfil\cr<\cr\noalign{\vskip1.5pt}\sim\cr}}}
{\vcenter{\offinterlineskip\halign{\hfil$\textstyle##$\hfil\cr<\cr
\noalign{\vskip1.0pt}\sim\cr}}}
{\vcenter{\offinterlineskip\halign{\hfil$\scriptstyle##$\hfil\cr<\cr
\noalign{\vskip0.5pt}\sim\cr}}}
{\vcenter{\offinterlineskip\halign{\hfil$\scriptscriptstyle##$\hfil
\cr<\cr\noalign{\vskip0.5pt}\sim\cr}}}}}

\begin{document}

\title{{\bf A dynamical and kinematical model of the Galactic stellar halo and
possible
implications for galaxy formation scenarios}}

\author{\\
J.~Sommer-Larsen$^{1}$, T.C.~Beers$^{5}$, C.~Flynn$^{3,4}$,\\
R.~Wilhelm$^{5}$ and P.~R.~Christensen$^{1,2}$\\
\\
1) Theoretical Astrophysics Center\\ Juliane Maries Vej 30, DK-2100 Copenhagen
{
\O},
Denmark\\
(jslarsen@tac.dk)\\
\\
2) The Niels Bohr Institute\\ Blegdamsvej 17, DK-2100 Copenhagen {\O},
Denmark\\
(perrex@nbi.dk)\\
\\
3) NORDITA\\Blegdamsvej 17, DK-2100 Copenhagen {\O}, Denmark\\
\\
4) Tuorla Observatory\\Piikki\"o, FIN-21500, Finland\\
(cflynn@astro.utu.fi)\\
\\
5) Department of Physics and Astronomy\\Michigan State University,
 E. Lansing Michigan 48824, USA\\
(beers@pa.msu.edu, wilhelm@pa.msu.edu)}

\date{ }
\maketitle
\vskip 1.0 truecm {\it Submitted to The Astrophysical Journal}

\newpage
\section*{Abstract}

We re-analyse the kinematics and dynamics of the system of blue
horizontal branch field (BHBF) stars in the Galactic halo (in
particular the outer halo), fitting the kinematics with the model of
radial and tangential velocity dispersions in the halo as a function
of galactocentric distance $r$ proposed by Sommer-Larsen, Flynn \&
Christensen (1994), but making use of a much larger sample of BHBF
stars than was previously available. The present sample consists of
nearly 700 Galactic halo BHBF stars.

A very good fit to the observations is obtained.  The basic result is
that the radial component, $\sigma_r$, of the stellar halo's velocity
ellipsoid decreases fairly rapidly beyond the solar circle.  The
observed decrease is from $\sigma_r \simeq 140\pm 10$ \kms at the sun,
to an asymptotic value of $\sigma_r = 89\pm 19$ \kms at large $r$.
The rapid decrease in $\sigma_r$ is matched by an increase in the
tangential velocity dispersion, $\sigma_t$, with increasing $r$.
Thus, the character of the stellar halo velocity ellipsoid is shown to
change markedly from radial anisotropy at the sun to tangential
anisotropy in the outer parts of the Galactic halo ($r\ga 20$ kpc).

The implications of our results for possible Galactic formation
scenarios are discussed. Our results may indicate that the Galaxy
formed hierarchically (partly or fully) through merging of smaller
subsystems - the 'bottom-up' galaxy formation scenario, which for
quite a while has been favoured by most theorists and recently also
has been given some observational credibility by HST observations of a
potential group of small galaxies, at high redshift, possibly in the
process of merging to a larger galaxy.

\vskip 1.7 truecm {\it Subject headings:} ~Galaxy: halo -- kinematics and
dynamics --
formation -- stars: horizontal-branch -- kinematics

\newpage

\section{Introduction}

Although the stellar halo accounts for only about 1\% of the luminous
mass of the Galaxy, it plays an crucial role in studies of the
Galaxy's formation, evolution, and present-day structure.  The halo
has long been considered the Galaxy's oldest component, age estimates
being tractable for its most conspicuous constituent, the metal-weak
globular clusters, as well as for (with less certainty) individual
metal-weak stars.  Thus the dynamical and chemical state of the
luminous halo population provides information on the formation of
large disk galaxies such as the Milky Way.  Furthermore, luminous
halo-population objects can be treated as dynamical tracers in order
to estimate the mass of the Galaxy's dark-matter halo, which
increasingly dominates the Galaxy's mass at large galactocentric
distances.

At present, the best measurements of the halo's kinematics are
obtained from analysis of the motions of stars found in the solar
neighbourhood; traditionally from samples of high-proper-motion stars
(Ryan \& Norris 1991; Carney et al. 1994), more recently from
kinematically-unbiased samples of metal-weak stars (see Beers \&
Sommer-Larsen 1995, and references therein).  In order to obtain
samples of halo-population stars {\it in situ}, methods have been
developed to find distant weak-lined K giants (Ratnatunga
\& Freeman 1989; Morrison, Flynn, \& Freeman 1990), RR Lyrae stars
(e.g. Hawkins 1984) and blue horizontal-branch
field (hereafter, BHBF) stars (Flynn, Sommer-Larsen \& Christensen 1994, 1995,
and references therein; Kinman, Suntzeff, \& Kraft 1994; Wilhelm, Beers, \&
Gray 1997).  Particular advantages of BHBF stars include the fact that they
are much easier to confidently identify than metal-weak K giants and
also considerably easier to identify than RR Lyrae stars.  BHBF
stars are also much more numerous than, for example, the RR Lyraes, and their
line-of-sight velocities can be determined comparatively much easier and
with higher accuracy.

In this paper we analyse the kinematics of a large sample of BHBF
stars in the outer halo.  We have supplemented the sample of 133 stars
used in a recent analysis of the kinematics and dynamics of the system
of BHBF stars of the outer Galactic halo by Sommer-Larsen, Flynn \&
Christensen (1994, hereafter SLFC94) with an additional 546 stars (for
a total of 679 BHBF stars), the majority of which were originally
identified in the HK survey of Beers and colleagues (Beers et
al. 1996, and references therein), and analysed using techniques
developed by Wilhelm (1995).  The data are described in section 2.  In
section 3 we fit the simple model, proposed by SLFC94 for the
dependence of the radial and tangential velocity dispersions of the
halo as a function of galactocentric distance $r$, to the enlarged
data sample.  We find that the radial component of the velocity
ellipsoid of the outer stellar halo is remarkably low, compared to the
value at the sun of $\sigma_{r,\odot} \simeq 140$ \kms, reaching an
asymptotic value of $\sigma_r \simeq 89$ \kms at large $r$, and that
the velocity ellipsoid in the outer halo is tangentially anisotropic,
in contrast to the local halo velocity ellipsoid, which is radially
anisotropic.  A discussion of the results obtained and their
implications, in particular for galaxy formation scenarios, is
presented in section 4. Finally in section 5 we summarize our
conclusions.

\section{The data}

The SLFC94 model is constrained by a specification of the halo
velocity ellipsoid in the vicinity of the sun, and line-of-sight
velocity measurements of halo BHBF stars in various fields over a
range of galactocentric distances from approximately 7 kpc to 65 kpc.

In the solar neighborhood the components of the velocity ellipsoid in
the halo are well known.  Norris, Bessell \& Pickles (1985) find
$\vec{\sigma} = (\sigma_r, \sigma_\phi, \sigma_\theta) = (125\pm 11,
96\pm 9, 88\pm 7)$
\kms for halo objects defined as [Fe/H] $ < -1.2$. Morrison, Flynn
\& Freeman (1990) found that the kinematics of the halo and thick disk
overlap in the range $-1.6 <$ [Fe/H] $< -1.0$.  In an attempt to isolate the
'true' halo, these authors used only stars with [Fe/H] $ < -1.6$ to derive
$\vec{\sigma} = (133\pm8, 98\pm13, 94\pm6)$ \kms. Beers \& Sommer-Larsen
(1995) also found significant evidence for an extension of the thick disk
well into the metallicity range usually identified with the Galactic halo.
After correction for the effects of thick-disk stars in the solar
neighbourhood, they derived a halo velocity ellipsoid of ~$\vec{\sigma} =
(153\pm
10, 93\pm 18, 107\pm 7)$ \kms. Considering the above, the radial velocity
dispersion at the sun is about $140$ \kms and the tangential components
($\sigma_\phi$ and $\sigma_\theta$) are about 90 - 100 \kms.  As is well known,
the local halo velocity ellipsoid is consequently quite radially anisotropic.

SLFC94 analysed a sample of 133 outer halo BHBF stars mainly located
in four fields: GP (NGP/SGP), $(l,b)=(0^\circ,\pm 90^\circ)$, F117,
$(l,b)=(270^\circ,-45^\circ)$, SA287, $(l,b)=(0^\circ,-47^\circ)$ and
22HR, (l,b)=$(38^\circ,-51^\circ)$.  Over the past few years, we have
obtained followup spectroscopy and broadband photometry for a sample
of some 1000 field horizontal-branch and other A-type stars identified
in the HK survey of Beers and collaborators.  Wilhelm (1995), and
Wilhelm, Beers, \& Gray (1997) discuss techniques which are suitable
for isolating a relatively pure sample of BHBF stars from this data
set.  A paper describing the full data set is presently in preparation
(Wilhelm et al. 1997).  For the purposes of the present analysis, we
have combined available data from the the Wilhelm et al.  catalog with
previously-published BFHB stars (including stars identified as FHB
from Norris 1986; Arnold \& Gilmore 1992; Beers, Preston, \& Shectman
1992; and Kinman, Sunzteff, \& Kraft 1994), obtaining a total of 546
additional BHBF stars.  The additional BHBF stars are located (within
a radius of $10^\circ$) in the GP fields, the SA287 field (plus one
symmetric reflection - at $(l,b)=(0^\circ,47^\circ)$, and the F117 and
22HR fields (plus three symmetric reflections - at
$(l,b)=(270^\circ,45^\circ),(90^\circ,-45^\circ), (90^\circ,45^\circ)$
and $(38^\circ,51^\circ),(322^\circ,-51^\circ),(322^\circ,51^\circ)$,
respectively).  The addition of stars in the 'reflection' fields
obviously requires an assumption that the halo is axially symmetric
and symmetric about the Galactic disk plane. We also assume that the
mean rotation of the halo is essentially zero (as found by, e.g.,
Sommer-Larsen \& Christensen 1989 and Beers \& Sommer-Larsen 1995).
Mild violation of these assumptions will not have serious impact on
our analysis but one should note that there are some indications of
kinematic substructure in the halo (Majewski et al. 1996 and
references therein).

In three of the four fields, including reflections, (GP, SA287 and
22HR), we have divided the stars into four bins, defined so that each
bin covers a well-defined range in galactocentric radius $r$.  In the
field F117, including reflections, we have fewer stars than in the
three other fields and have only used three bins. For each bin, we
have calculated the line-of-sight velocity dispersion $\sigma_{\rm
los}$. The results are given in Table 1. The mean distance from the
sun $<d>$, the mean galactocentric distance $<r>$, the mean value of
the geometric projection factor $<\gamma>$ (equation 6 below) and the
number of stars $N$ in each bin are also listed.

Recalling that for larger $r$, the line-of-sight velocity dispersion
$\sigma_{\rm los}$ is essentially a measure of $\sigma_r$, the clear
implication of Table 1 is that the radial velocity dispersion of the
outer halo is significantly lower than that found in the solar
neighbourhood. In the most distant bin in F117, $\sigma_{\rm los} =
107\pm24$ \kms at $r \simeq 21$ kpc, in 22$^{h}$, $\sigma_{\rm los} =
112\pm12$ \kms at $r \simeq 16$ kpc, in SA287, $\sigma_{\rm los} =
113\pm20$
\kms at $r \simeq 13$ kpc, and  at the GP, $\sigma_{\rm los} = 99\pm19$ \kms
at $r \simeq 19$ kpc.  For the very distant stars ($r \ga 45$ kpc), the
measured
line-of-sight velocity dispersion is $100\pm 23$ \kms at a mean galactocentric
distance of $54\pm2$ kpc. In summary, the data suggest that $\sigma_r$
decreases
from about 140 \kms locally to about 90 -110 \kms at large $r$.

\section{The model fit}

The model is described in detail in SLFC94 and is briefly
recapitulated here to make this paper more easily readable.

It is generally found that for $r \ga r_{\odot}$, where $r_\odot$ =
8.0 kpc is the solar galactocentric distance, the stellar halo is
approximately spherical - see, e.g., the review by Freeman (1987) and
references therein; Hartwick (1987); Sommer-Larsen \& Zhen (1990);
Yamagata \& Yoshii (1992), but note that there are indications that
the inner halo is somewhat flattened (e.g. Hartwick 1987; Larsen \&
Humphries 1994; Wetterer \& McGraw 1996). In this paper we shall be
concerned with the properties of the outer Galactic halo and
consequently the outer stellar halo (as traced by BHBF stars) is
assumed to be approximately spherical. Furthermore it will be assumed
that the density fall-off of the BHBF stars of outer stellar halo can
be approximated by the power-law relation $\rho (r) \propto
r^{-\alpha}, ~\alpha=3.4 \pm 0.3$ - see SLFC94.

The potential of the outer parts of the Galaxy is assumed to be
approximately spherical and logarithmic, $\Phi (r) = V_{c}^{2}$ ln
$(r)$, corresponding to a flat rotation curve of the Galaxy, with
$v_c(r) = V_c$ = 220
\kms.

The anisotropy parameter $\beta$ is defined as $$ \beta = 1 -
(\frac{\sigma_t}{\sigma_r})^2\;, \eqno(1) $$ where $\sigma_r$ is the
radial and $\sigma_t$ the (1-D) tangential velocity dispersion.  Thus
the Jeans equation has the form $$ \frac{1}{\rho}
\frac{d(\rho\sigma_r^2)}{dr} +
\frac{2\beta\sigma_r^2}{r} = - \frac{d\Phi}{dr}.  \eqno(2) $$

In section 2, the evidence pointed to a significant decrease in the
radial velocity dispersion, from about 140 \kms at the sun to about 90
-- 110 \kms in the outer halo. We use the following simple model for
$\sigma_r^2(r)$: $$
\sigma_r^2 = \sigma_0^2+\frac{\sigma_+^2}{\pi}(\frac{\pi}{2}-{\rm
tan}^{-1}(\frac{r-r_0}{l})).\eqno(3)$$ Adopting this form gives us good
flexibility in modelling the decrease in $\sigma_r(r)$ with increasing $r$.

It follows from equation (3) that $\sigma_0$ is the asymptotic value
of the radial velocity dispersion for $r >> (r_0 + l)$ and that
$\sqrt{\sigma_+^2 + \sigma_0^2}$ approximately is the radial velocity
dispersion in the inner halo ($r \la r_{\odot}$). The physical meaning
of the two scale parameters $r_0$ and $l$ is given in SLFC94 and is
fairly straightforward.

Substituting (3) into (2), we solve for $\sigma_t$: $$ 2\sigma_t^2 =
V_{\rm c}^2-\sigma_r^2(\alpha-2) -\frac{1}{\pi}\frac{r}{l}
\frac{\sigma_+^2}{(1+((r-r_0)/l)^2)}. \eqno(4)$$

The line-of-sight velocity dispersion of a set of stars in a field at
Galactic coordinates $(l,b)$ where the velocity ellipsoid has
components $\sigma_r$ and $\sigma_t$ is $$\sigma_{\rm los}^2 =
\gamma^2\sigma_r^2+(1-\gamma^2)\sigma_t^2\;, \eqno(5)$$ where $\gamma$ is a
simple geometric projection factor given by $$ \gamma = (d - {r_\odot}{\rm
cos}\,l\,{\rm cos}\,b)/r. \eqno(6)$$ The mean values of $\gamma$ for the
various bins are given in Table 1.

We have fitted the above model to the much larger data sample
considered in this paper using a maximum likelihood approach, assuming
a radial velocity dispersion at the sun, $r=r_\odot$, of
$\sigma_{r,\odot}$ = 140 \kms, as in SLFC94, and hence reducing the
number of free parameters of the model to three.

The best-fit model is shown in Figure 1, where we show $\sigma_r(r)$
as a solid line and $\sigma_t(r)$ as a dashed line. The best-fit
parameters are $r_0=13.5$ kpc, $l = 7.5$ kpc, $\sigma_0 = 89$ \kms and
$\sigma_+$ = 129 \kms.  The radial velocity dispersion decreases from
148 \kms in the inner halo to an asymptotic value $89$ \kms at large
$r$, whereas the tangential velocity dispersion increases from about
91 \kms in the inner halo to about 137 \kms in the outer halo.  The
radial velocity dispersion, $\sigma_r$, decreases fairly rapidly
beyond the solar circle, with a corresponding increase in $\sigma_t$.
The model predicts a tangential velocity dispersion at the sun,
$r=r_\odot$, of $\sigma_{t,\odot}$ = 93 \kms, in good agreement with
the measured value for this quantity (see section 2), which is about
90 -100 \kms.

In Figure 2 $\sigma_{\rm los}$, as predicted by the model, is shown as
the solid lines for the fields described in section 2. The circles are
the data, the vertical error bars being the observational 1-$\sigma$
error in the determination of $\sigma_{\rm los}$ and the horizontal
bars the 1-$\sigma$ error in the mean distance for the stars in the
various bins.  The value of $\chi^2$ for the model is 8.0 (13 degrees
of freedom), a very good fit, as can also be seen from inspection of
Figure 2.

\section{Discussion}

\subsection{The dynamics and kinematics of the outer stellar halo}

The main result of the analysis is that $\sigma_r$ decreases fairly
rapidly beyond the solar circle -- already by $r \simeq 20$ kpc it has
dropped to approximately 110 \kms; by $r \simeq 40$ kpc it has
decreased to about 90 \kms.  This conclusion is very firm because the
radial velocities in our outer fields are dominated by $\sigma_r$.
The tangential velocity dispersion, $\sigma_t$, is expected to rise
correspondingly rapidly in this region in order to be consistent with
a flat rotation curve.  We note here that the decrease in $\sigma_r$
and increase in $\sigma_t$ actually takes place within $r \la 20$ kpc,
where the rotation curve is observationally well constrained to be
flat (Fich \& Tremaine 1991).

We find that the 1-$\sigma$ error on $\sigma_0$ is $\pm 19$ \kms, {\it
i.e}, $\sigma_r$ decreases from $140 \pm 10$ \kms at the sun to about
$89\pm 19$ \kms at large $r$.  Hence, the most important kinematic
feature of the model is that the velocity ellipsoid changes from {\it
radial} anisotropy in the solar vicinity ($\beta \simeq$ 0.5) to {\it
tangential} anisotropy in the outer halo ($\beta \simeq$ -1.3).

To illustrate this further, we have calculated the predictions of two,
alternative 'toy' models.  For these models we assumed that the
velocity ellipsoid is constant with $r$, rather than depending quite
strongly on $r$, as for the dynamical model presented in SLFC94 and in
this paper.  For the first 'toy' model ~$\vec{\sigma}$ = (150,100,100)
\kms (in spherical polars) was assumed, as indicated by the findings
of Beers \& Sommer-Larsen (1995). The second 'toy' model was identical
to the first, except that it had $\sigma_r$ = 140 \kms, since this
value of $\sigma_r$, at $r=r_\odot$, was adopted in our main model, as
described previously in this section.

We compared the predictions of the 'toy' models with the data by
calculating the value of $\chi^2$ obtained. For the first 'toy' model
$\chi^2$ = 46.5 (13 degrees of freedom) implying that it can be
rejected at the 99.999 \% confidence level. For the second $\chi^2$ =
33.8 (13 degrees of freedom), so this model can 'only' be rejected
with 99.9 \% confidence. Clearly, such models are not viable and this,
of course, is the main reason why the dynamical model of Sommer-Larsen
(1987), where $\vec{\sigma}$ has a rather strong dependence on $r$,
and all subsequent similar models (including the one discussed in this
paper), has been proposed.

\subsection{Outer stellar halo kinematics: clues towards understanding
the formation of the Milky Way}

Our results concerning the dynamics and kinematics of the outer
stellar halo are of considerable interest in relation to theories of
the formation of the Milky Way, in particular, and galaxies in
general.

If the Galaxy formed from a single collapsing over-density region in
the early universe, then one might expect the outer halo to be
characterized by significantly-radially-anisotropic kinematics (see,
e.g., van Albada 1982), whereas the data show that quite the opposite
is the case. If, on the other hand, at least the outer parts of the
proto-Galaxy were assembled by accretion of various lumps, then a
large tangential velocity dispersion in the outer parts of the Galaxy
is possible, depending on the nature of the accretion (Norris 1994;
Freeman 1996).  So our results indicate that the outer stellar halo
formed by some sort of accretion and merging processes. The kinematics
of stars in the inner halo are, at least locally, radially
anisotropic, possibly indicating that the inner parts of the halo
formed during a more dissipative and coherent collapse, probably on a
relatively short (dynamical) time scale and possibly in concert with
ongoing accretion and merging processes in the inner halo as well.

Chemical evolution arguments lead to a similar conclusion on the basis
of the finding that there is a significant abundance gradient in the
inner halo, but essentially none in the outer halo. This was discussed
in the pioneering work by Searle \& Zinn (1978) and in much subsequent
work - see, e.g., Norris (1996) and references therein.
\\
\\
In the following we discuss various aspects of the formation of the Galaxy in
more detail.

It is possible that the major part of the inner gas ended up in the
Galactic bulge as the bulge and the local stellar halo have very
similar specific angular momentum distributions (Wyse \& Gilmore 1992;
Ibata \& Gilmore 1995).

The formation of the outer stellar halo may, on the one hand, not
necessarily be related to the formation of the Galactic disk, since
the thick disk apparently contains stars of very low metallicity --
comparable to or perhaps lower than the average abundance of the
globular clusters (Beers \& Sommer-Larsen 1995).  On the other hand,
the globular clusters are not likely to be representative of the
stellar halo (see below) and it seems somewhat more direct and
reasonable to assume that the Galactic disk mainly {\it did} form out
of gas, initially located in subsystems gradually merging in the inner
part of the increasingly deep, dark-matter potential well. In the
following we shall consider this latter and, perhaps, most relevant
option in more detail.

It is likely that the Galactic globular clusters survived to the
present only because they are so compact, whereas the more diffuse and
probably lower-metallicity subsystems, which contained the major part
of the gas available for the formation of the Galactic disk (see
below), broke up quite early due to effects of star-formation feedback
processes, tidal destruction, dynamical friction etc. (see also
Freeman 1996). The bulk of gas in these systems most likely was left
in a dilute, non star-forming, state after the disruption and
subsequently gradually settled as a large and massive, differentially
rotating disk.

The mass of the Galactic globular cluster system is only of order 0.2
\% of the mass of the Galactic disk, so the globular clusters are
likely to be highly unrepresentative of the typical accreted
subsystems.

Furthermore the mass of the stellar halo is of the order 2 \% of the
mass of disk, indicating that only a tiny fraction of the gas in the
more typical subsystems was locked up as stars prior to the disruption
of the subsystems.

The famous, solar neighbourhood, G-dwarf problem is most readily
resolved if it is assumed that gas continued to settle onto the
Galactic disk over a period, which is shorter than, but comparable to,
the age of the disk (e.g. Pagel \& Patchett 1975; Sommer-Larsen 1991;
Rocha-Pinto \& Maciel 1996).  If this indeed is the correct resolution
of the G-dwarf problem, then the settling of the gas from the
disrupted subsystems onto the Galactic disk was quite gradual, as
proposed above.

The key question seems to be: From where did the large amount of
angular momentum of the Galactic disk originate ?

It follows from the work of Sommer-Larsen \& Zhen (1990) that the
time-averaged galactocentric distance of local halo stars is, in
general, smaller than $r_{\odot}$. The average rotation velocity of
the local halo stars is very small, perhaps even negative,
(e.g. Sommer-Larsen \& Christensen 1989; Beers \& Sommer-Larsen 1995)
and hence so is their average specific angular momentum. Consequently,
the gas associated with the formation of the local halo stars, at $r
\la r_{\odot}$, most likely did not end up in the Galactic disk.
Indeed, as argued above, this gas probably ended up in the bulge.

So, in this scenario, the bulk of the disk gas originated from the
disrupted subsystems that formed the outer stellar halo.  Consequently
these systems, taken as a whole, were carrying an amount of angular
momentum, probably mainly in orbital form, which was at least as large
as that of the present Galactic disk. The following simplistic
discussion indicates that this prediction can be tested
observationally.

The surface density distribution
of a truncated, exponential disk, as a function\\
\\
of radial coordinate $R$, is given by
\setcounter{equation}{6}
\begin{eqnarray}
                                        & \Sigma_0 \exp (-R/R_d)~ , & ~~R < R_t
\nonumber \\
\Sigma (R)  =  \Bigg\{ & &     \\
                                         & 0 ~~~~~~~~~~~~~~~~~~~ , & ~~R \ge
R_t \nonumber
\end{eqnarray}
where $\Sigma_0$ is the central surface density, $R_d$ the exponential
scale length and $R_t$ the truncation radius. The stellar component of large
galactic disks is typically truncated
at $R_t \sim 4 R_d$.  Since the mass of the stars in the Galactic disk
is much larger than that of the gas, we neglect, for simplicity,
the latter component in this discussion.
Assuming a constant rotation curve, i.e. $v_c (R) \simeq V_c = c^{st}$, it
easy to show that the
specific angular momentum of the disk is
\begin{equation}
j_d = \xi (q) R_d V_c ~~, ~q = \frac{R_t}{R_d} ~~,
\end{equation}
where $\xi(q)$ = 1.44, 1.68 and 1.82 for $q$ = 3, 4 and 5 respectively
(and $\xi (q) \rightarrow 2$ for $q \rightarrow \infty$).

Let the typical galactocentric distance of disruption of the Galactic disk
progenitor
systems be denoted $r_{disrupt}$. The results obtained in this paper suggest
that the kinematics
of the stars of the outer halo are quite tangentially anisotropic
($\beta \sim
-1.3$) and hence that the average galactocentric distance of the halo stars
originating from the disk progenitor systems is approximately equal
to $r_{disrupt}$. The mean specific angular momentum of the dilute gas,
originating from the disrupted subsystems and later settling onto the
disk, is likely to be approximately conserved during this infall phase
(contrary to the case where
the gas is spiraling inwards as dense satellites, continuously losing
orbital angular momentum and energy to the dark matter halo by dynamical
friction - see, e.g., Navarro \& White 1994). Denoting the mean
rotational velocity of the halo stars at $r \sim r_{disrupt}$ by
$\bar{v}_{rot} = <v_{\phi}>$ it then follows that
\begin{equation}
j_d \sim r_{disrupt} ~\bar{v}_{rot} ~~,
\end{equation}
so
\begin{equation}
\bar{v}_{rot} \sim \frac{\xi(q) R_d V_c}{r_{disrupt}} =
49 ~(\frac{\beta(q)}{1.68}) ~(\frac{R_d}{4 ~{\rm kpc}}) ~(\frac{V_c}{220 ~{\rm
km/s}})
~(\frac{r_{disrupt}}{30 ~{\rm kpc}})^{-1} ~{\rm km/s} ~.
\end{equation}

To briefly recapitulate the essentials of the observational test: If
$\bar{v}_{rot} \ga$ 50 \kms in the outer halo, then it appears quite
likely that the subsystems, from where the stars in the outer halo
originated, were the main progenitors of the Galactic disk.
Conversely, if $\bar{v}_{rot}$ is much smaller than this, perhaps even
negative, in the outer halo (like for the local halo stars), then it
seems unlikely that the progenitor subsystems of the outer stellar
halo were also the the main progenitors of the Galactic disk.

Unfortunately, it is doubtful that this prediction can be tested on
the basis of observed line-of-sight velocities of outer halo BHBF
stars, since, as discussed previously, the line-of-sight velocity is
essentially the radial velocity, $v_r$, at large $r$. To determine
$\bar{v}_{rot}$ observationally, on the basis of proper motions,
requires a measurement accuracy of about $10^{-4}$ "/yr, so this is
clearly not possible on the basis of photographic plates, even with a
~$\sim$ 100 yr baseline. Nor can it be done using the Hipparcos
satellite, as the stars are quite faint ($B \simeq V \sim 18$ mag),
and as the measuring accuracy of even Hipparcos is an order of
magnitude too low for this purpose.  But the discussion above suggests
that it would be of considerable relevance and importance to bear in
mind this observational test, based on outer halo stars, when planning
future proper motion measuring systems.

All in all our results may be interpreted as supporting the notion
that the Galaxy formed hierarchically (partly or fully), preying on
smaller subsystems - the 'bottom-up' galaxy formation scenario (see,
e.g., Blumenthal et al. 1984 for an excellent review of various galaxy
formation scenarios), which for quite a while has been considered the
favorite by the majority of theorists working on galaxy formation and
recently also has been given some observational credibility through
HST observations of a potential group of small galaxies, at high
redshift, possibly in the process of merging to a larger galaxy
(Pascarelle et al. 1996).

\section{Conclusion}

We have analysed the kinematics and dynamics of the system of blue horizontal
branch field (BHBF) stars in the outer Galactic halo using a large sample
($\sim 700$) of BHBF stars.

The basic result is that the radial component, $\sigma_r$, of the
stellar halo's velocity ellipsoid decreases fairly rapidly beyond the
solar circle, from $\sigma_r \simeq 140 \pm 10$ \kms, at $r =
r_\odot$, to an asymptotic value of $\sigma_r = 89\pm 19$ \kms at
large $r$.  This result is very firm, because at large $r$,
$\sigma_{\rm los}$ is dominated by $\sigma_r$.  Assuming that the
rotation curve of the Galaxy is approximately flat, the fairly fast
decrease in $\sigma_r$ is matched by an increase in the tangential
velocity dispersion, $\sigma_t$ with increasing $r$.  We conclude that
the stellar halo velocity ellipsoid changes markedly, from radial
anisotropy ($\beta \simeq 0.5$) at the sun, to tangential anisotropy
($\beta
\simeq -1.3$) in the outer parts of the Galactic halo ($r\ga 20$ kpc).

The implications of our results for possible Galactic formation
scenarios are discussed. Our results may indicate that the Galaxy
formed hierarchically (partly or fully) through merging of smaller
subsystems - the 'bottom-up' galaxy formation scenario, which for
quite a while has been favoured by most theorists and recently also
has been given some observational credibility by HST observations of a
potential group of small galaxies, at high redshift, possibly in the
process of merging to a larger galaxy (Pascarelle et al. 1996).

The results obtained in this paper are very similar to what SLFC94
found based on a five times smaller sample of BHBF stars. It is
gratifying, though, that the goodness of the model fit to the
observational data has improved considerably relative to model fit in
SLFC94.

The recent work of Flynn, Sommer-Larsen \& Christensen (1996) strongly
suggests, on the basis of numerical simulations, that such a model is
physically feasible, in the sense that a stationary phase space
distribution function $f$ exists, which can generate this sort of
kinematics and at the same time is non-negative everywhere in phase
space. This is also strongly hinted at by the theoretical, dynamical
models of Sommer-Larsen (1987) and Vedel \& Sommer-Larsen (1990).

\section*{Acknowledgements}

We have benefited considerably from the comments of Bernard Pagel,
Henrik Vedel, Draza Markovi\'{c}, Cedric Lacey and the referee.  This
work was supported by Dansk Grundforskningsfond through its support
for an establishment of the Theoretical Astrophysics Center.
T.C.B. acknowledges partial support for this work from grants AST
90-1376 and AST 92-22326 awarded by the National Science Foundation.
T.C.B. would like to express gratitude for the hospitality shown by
the Theoretical Astrophysics Center at the University of Copenhagen,
where the work reported in this paper was conducted.  R.W.
acknowledges partial support from the teaching post-doctoral program
at Michigan State University.

\newpage
\vskip 5 truecm
\begin{table}[p]
\small
\caption{Line-of-sight velocity dispersions etc. for the 16 data bins}
\begin{center}
\begin{tabular}{l|rcr|ccccc}
\hline\hline
Field         &     & $d$ [kpc]    &       &  $<d>$ [kpc]     &$<r>$ [kpc]
   &$\sigma_{\rm los}$
[\kms] & $<\gamma>$  & $N$   \\
\hline
F117          &$   0<$&$d$&$<~6$  &$ ~4.0\pm 0.3 $&$ ~8.9$&$  120\pm15$ & ~0.45
 &  39 \\
              &$6\le$&$d$&$<12$  &$ ~8.0\pm 0.4 $&$ 11.3$&$  113\pm10$ & ~0.71
&  33 \\
              &$12\le$&$d$&$<45$  &$ 19.0\pm 1.6 $&$ 20.6$&$  107\pm24$ & ~0.92
 &  21
\\
\hline
22HR          &$   0<$&$d$&$<~4$  &$  ~3.4\pm 0.2 $&$ ~7.0$&$  ~87\pm10$ &
-0.01
  & 44 \\
              &$4\le$&$d$&$<~6$  &$ ~5.0\pm 0.1 $&$ ~7.0$&$  111\pm10$ & ~0.15
& 57 \\
              &$6\le$&$d$&$<12$  &$ ~7.8\pm 0.2 $&$ ~7.5$&$  116\pm15$ & ~0.48
& 47 \\
              &$12\le$&$d$&$<45$  &$ 18.7\pm 1.3 $&$ 16.3$&$  112\pm12$ & ~0.90
 & 29 \\
\hline
SA287         &$   0<$&$d$&$<~4$  &$  ~3.0\pm 0.2 $&$ ~6.3$&$ ~93\pm11$ & -0.15
 & 52 \\
              &$4\le$&$d$&$<~6$  &$ ~5.0\pm 0.2 $&$ ~5.9$&$ ~97\pm~9$ & -0.01
 &
  76 \\
              &$6\le$&$d$&$<12$  &$ ~7.6\pm 0.2 $&$ $~6.2&$ ~93\pm13$ & ~0.34
 &
  66 \\
              &$12\le$&$d$&$<45$  &$ 16.7\pm 0.9 $&$12.7$&$ 113\pm20$ & ~0.89
 &
  28 \\
\hline
GP           &$   0<$&$d$&$<~4 $  &$  ~2.8\pm 0.1 $&$ ~8.5$&$  ~77\pm~8$ & ~0.
33  & 52 \\
              &$ 4\le$&$d$&$<~6$  &$ ~4.8\pm 0.1 $&$ ~9.3$&$  ~95\pm10$ & ~0.51
 & 66 \\
              &$ 6\le$&$d$&$<12$  &$ ~7.4\pm 0.2 $&$ 10.9$&$  117\pm12$ & ~0.68
 & 43 \\
              &$ 12\le$&$d$&$<45$  &$ 17.6\pm 2.7 $&$ 19.3$&$  ~99\pm19$ &
~0.91
  & 21 \\
\hline
Very distant  &$   $&$d$&$\ge45$  &$ 55.2\pm 2.4 $&$ 53.9$&$ 100\pm23$ & ~0.99
&  ~9 \\
\hline \hline
\end{tabular}
\end{center}
\end{table}

\newpage

\section*{References}
\begin{trivlist}

\item[] Arnold, R., \& Gilmore, G. 1992, MNRAS, 257, 225
\item[] Beers, T.C.,  \& Sommer-Larsen, J. 1995, ApJS, 96, 175
\item[] Beers, T.C., Preston, G.W.,  \& Shectman, S.A. 1992, AJ, 103, 1987
\item[] Beers, T.C., Doinidis, S.P., Wilhelm, R.J.,  \& Mattson, C. 1996,
        ApJS, 103, 433
\item[] Blumenthal, G.R., Faber, S.M., Primack, J.R., \& Rees, M.J. 1984,
Nature,
\item[] \hspace{1.0cm} 311, 517
\item[] Carney, B.W., Latham, D.W., Laird, J.B., \& Aguilar, L.A. 1994, AJ,
107, 2240
\item[] Fich, M.,  \& Tremaine, S. 1991, ARA\&A, 29, 409
\item[] Flynn, C., Sommer-Larsen, J.,  \& Christensen, P.R. 1994, MNRAS, 267,
77
\item[] Flynn, C., Sommer-Larsen, J.,  \& Christensen, P.R. 1995, A\&AS, 109,
171
\item[] Flynn, C., Sommer-Larsen, J.,  \& Christensen, P.R. 1996, MNRAS, 281,
1027
\item[] Freeman, K.C. 1987, ARA\&A, 25, 603
\item[] Freeman, K.C. 1996, in Formation of the Galactic Halo - Inside and Out,
ed.
\item[] \hspace{1.0cm} H. L. Morrison \& A. Sarajedini (Astronomical Society of
the Pacific:
\item[] \hspace{1.0cm} San Francisco), 3
\item[] Hartwick, F.D.A. 1987, in The Galaxy, ed. G. Gilmore \& R.Carswell
(Dordrecht:
\item[] \hspace{1.0cm} Reidel), 281
\item[] Hawkins, M.R.S. 1984, MNRAS, 206, 433
\item[] Ibata, R.,  \& Gilmore, G. 1995, MNRAS, 275, 605
\item[] Kinman, T.D., Suntzeff, N.B.,  \& Kraft, R.P. 1994, AJ, 108, 1722
\item[] Larsen, J.A. \& Humphries, R.M. 1994, ApJ, 436, L149
\item[] Majewski, S.R., Munn, J. A.,  \& Hawley, S. L. 1996, ApJ, 459, L73
\item[] Morrison, H., Flynn, C.,  \& Freeman, K.C. 1990, AJ 100, 1191
\item[] Navarro, J.F., \& White, S.D.M. 1994, MNRAS, 267, 401
\item[] Norris, J.E. 1986, ApJS, 61, 667
\item[] Norris, J.E. 1994, ApJ, 431, 645
\item[] Norris, J. E. 1996, in Formation of the Galactic Halo - Inside and Out,
ed.
\item[] \hspace{1.0cm} H. L. Morrison \& A. Sarajedini (Astronomical Society of
the Pacific:
\item[] \hspace{1.0cm} San Francisco), 14
\item[] Norris, J., Bessell, M.,  \& Pickles, A. 1985, ApJS, 58, 463
\item[] Pagel, B. E. J., \& Patchett, B. E. 1975, MNRAS, 172, 13
\item[] Pascarelle, S.M., Windhorst, R.A., Keel, W.C., \& Odewahn, S.C. 1996,
Nature,
\item[] \hspace{1.0cm} 383, 45
\item[] Ratnatunga, K.,  \& Freeman, K.C. 1989, ApJ, 339, 126
\item[] Rocha-Pinto, H. J., \& Maciel, W. J. 1996, MNRAS, 279, 447
\item[] Ryan, S.G.,  \& Norris, J.E. 1991, AJ, 101, 1835
\item[] Searle, L., \& Zinn, R. 1978, ApJ, 225, 357
\item[] Sommer-Larsen, J. 1987, MNRAS, 227, 21p
\item[] Sommer-Larsen, J. 1991, MNRAS, 249, 368
\item[] Sommer-Larsen, J.,  \& Christensen, P.R. 1989, MNRAS, 239, 441
\item[] Sommer-Larsen, J., Flynn, C.,  \& Christensen, P.R. 1994, MNRAS, 271,
94
\item[] Sommer-Larsen, J., \& Zhen, C. 1990, MNRAS, 242, 10
\item[] van Albada, T.S. 1982, MNRAS, 201, 939
\item[] Vedel, H.,  \& Sommer-Larsen, J. 1990, ApJ, 359, 104
\item[] Wilhelm, R. 1995, PhD thesis, Michigan State University
\item[] Wilhelm, R., Beers, T.C.,  \& Gray, R. 1997, AJ, submitted
\item[] Wilhelm, R., Beers, T.C., Flynn, C., Layden, A., Pier, J.R., \&
\item[] \hspace{1.0cm} Sommer-Larsen, J. 1997, in preparation
\item[] Wetterer, C.J., \& McGraw, J.T. 1996, AJ, 112, 1046
\item[] Wyse, R.F.G.,  \& Gilmore., G. 1992, AJ, 104, 144
\item[] Yamagata, T., \& Yoshii, Y. 1992, AJ, 103, 117
\end{trivlist}


\pagestyle{empty}

\newpage
\begin{figure}
{\bf Figure 1}  Best-fit model. The solid line shows the radial velocity
dispersion $\sigma_r(r)$ and the dashed line shows the tangential velocity
dispersion $\sigma_t(r)$.
\input epsf
\centering
\leavevmode
\epsfxsize=1.0
\columnwidth
\epsfbox{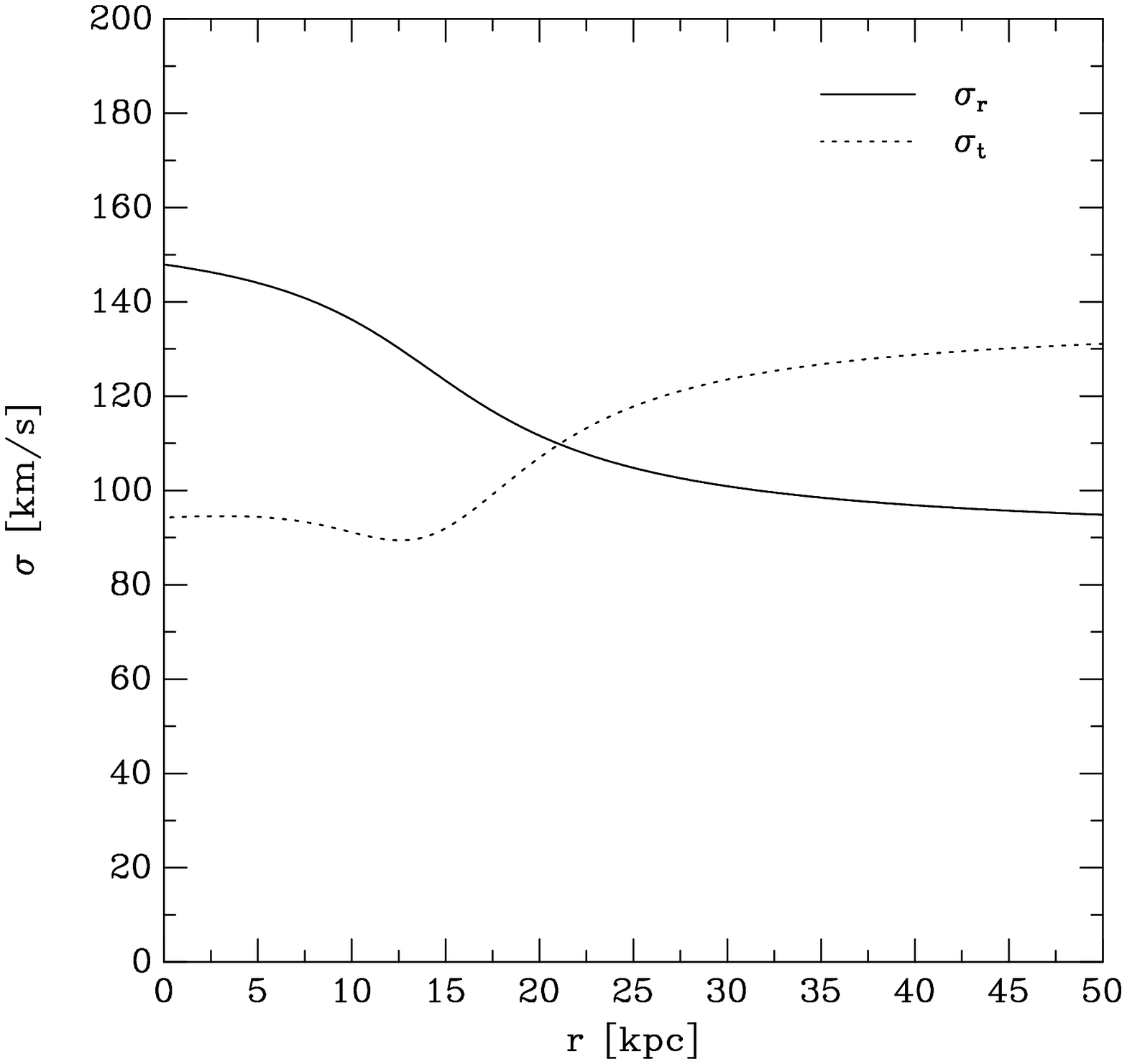}
\end{figure}

\newpage
\begin{figure}
{\bf Figure 2} The line-of-sight velocity dispersion for the best fitting
model as a function of distance $d$ along the four lines of sight
(including reflections) is shown as solid lines.  Solid circles show
the measured values from Table 1.  The vertical error bars are the
1-$\sigma$ errors in the line-of-sight velocity dispersion
measurement, and the horizontal error bars are the 1-$\sigma$ errors
in the mean distance of the stars in each bin.  The predictions of
'toy' model \#1, with ~$\vec{\sigma}$ = (150,100,100) \kms in
spherical polars, is shown by the dotted lines. Similarly, 'toy' model
\#2, with ~$\vec{\sigma}$ = (140,100,100) \kms in spherical polars, is
shown by the dashed lines.

\input epsf
\centering
\leavevmode
\epsfxsize=1.0
\columnwidth
\epsfbox{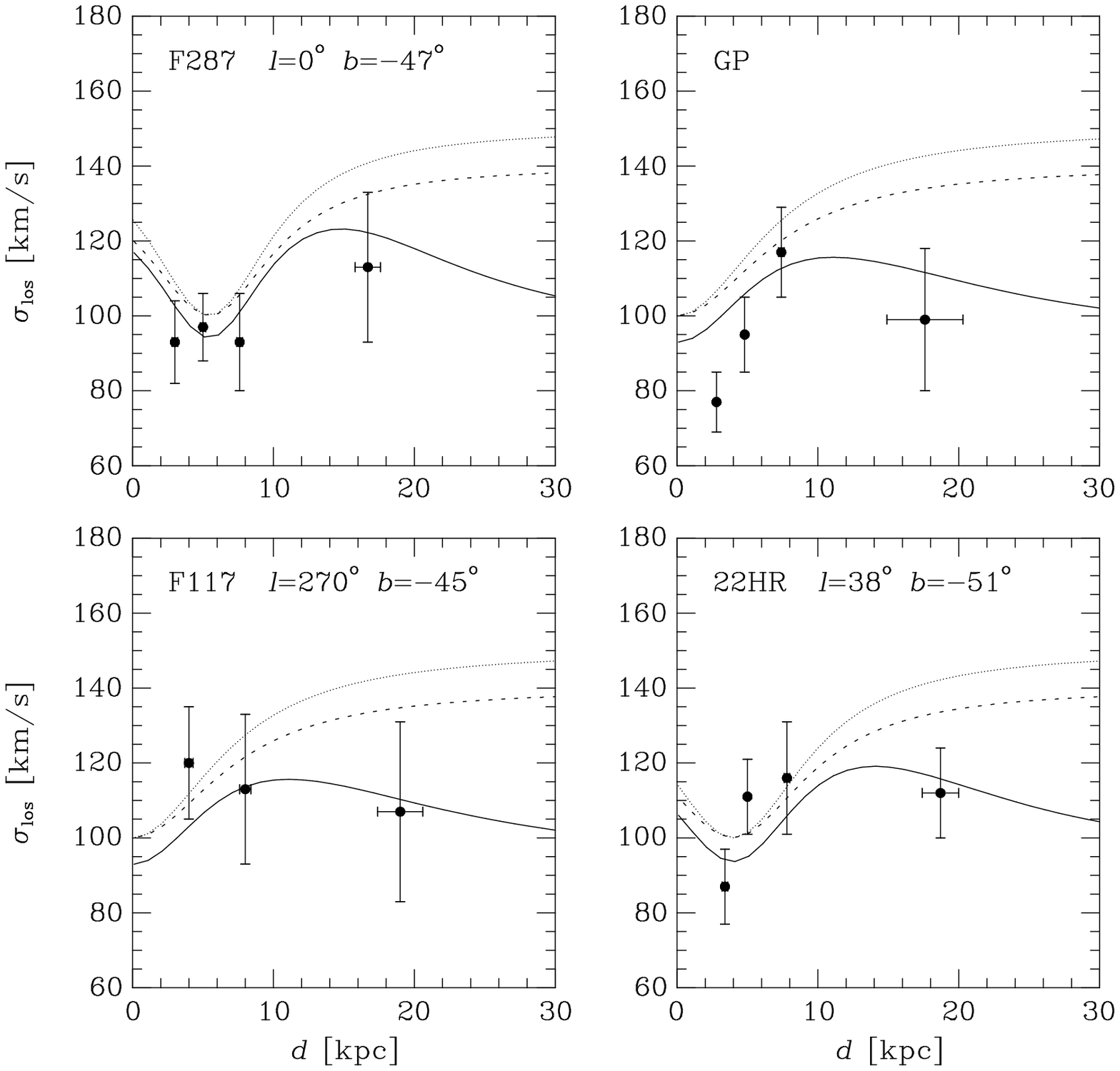}
\end{figure}

\end{document}